\documentclass[12pt]{article}
\usepackage{latexsym,graphicx,multirow}
\usepackage{float}
\usepackage{amssymb}
\usepackage{amsmath}
\usepackage{amscd}
\usepackage{amsthm}
\usepackage[left=2cm,top=2.5cm,right=2.5cm,bottom=1.5cm]{geometry}
\usepackage{hyperref}
\usepackage{epstopdf}
\usepackage{cite}
\usepackage{caption}
\usepackage{subcaption}
\usepackage{color}
\usepackage[utf8]{inputenc}
\usepackage[T1]{fontenc}
\begin{document}
	
\begin{center}
\large{\bf{Quintessence scalar field model in Weyl-type $f(Q,T)$ Gravity with $w_D-w'_D$ analysis }} \\
		\vspace{10mm}
\normalsize{Vinod Kumar Bhardwaj$^1$,  Priyanka Garg$^{2}$  }\\
		\vspace{5mm}
\normalsize{Department of Mathematics, GLA University, Mathura-281 406,\\ Uttar Pradesh, India}\\
		\vspace{2mm}
%\normalsize{$^{3}$Centre for Cosmology, Astrophysics and Space Science (CCASS), GLA University,\\	Mathura -281 406, Uttar Pradesh, India}\\
%		\vspace{2mm}
$^1$E-mail:dr.vinodbhardwaj@gmail.com \\
		\vspace{2mm}
$^2$E-mail:pri.19aug@gmail.com\\
		\vspace{2mm}

		%\date{}
		%\maketitle
\end{center}
\begin{abstract}
In the present study, we explore the dynamical characteristics of the quintessence cosmological model in Weyl-type $f(Q,T)$ gravity.  Here, $T$ represents the trace of the matter energy-momentum tensor, and $Q$ symbolizes the nonmetricity tensor. We propose a solution to the field equation using the specific parametrization in the form$H(z) = H_{0} (1+z)^{1+\alpha+\beta} e^{\left(\frac{- \beta z}{1+z}\right)}$, which depicts the necessary transition of cosmos from decelerating era to the current accelerating scenario. The values of model parameters are estimated as $H_0 = 71.17\pm 0.25 $, $\alpha = -0.663\pm0.030$, and $\beta = 1.488\pm0.087$ using the MCMC analysis and limiting the model with a joint dataset of Pantheon, BAO, and OHD. We discuss the cosmic behavior of many features of the derived model like EoS parameters, energy density, and cosmic pressure. Further, we have also explored the cosmological behavior of the quintessence model in Weyl $f(Q,T)$gravity. We have described the cosmic behavior of the model by $\omega_D-\omega_D'$ analysis. The diagnosis of the model is also performed using state finders and jerk parameters. In the end, we have discussed the energy conditions for the proposed model. Our analysis shows that the suggested model is well consistent with the recent findings.
\end{abstract}
\smallskip 
{\bf Keywords} :  Weyl-type $f(Q,T)$ gravity, Observational constraints, State finders, Quintessence DE model, $\omega_D-\omega_D'$ analysis.
	
PACS: 98.80.-k \\

\section{Introduction}
In the early 20th century, astronomers like Edwin Hubble suggested the idea of expanding the universe when they saw that galaxies were moving apart from one another.  But in the late 1990s, supernova experiments added a new unexpected finding in the expansion discovery of the universe which is that the universe was not only expanding, but it was growing faster. Later on, numerous cosmological observations have also confirmed these cosmological findings \cite{ref1,ref2,ref3,ref4,ref5,ref6,ref7}. To explain this accelerated expansion, scientists proposed the existence of a mysterious form of energy called Dark energy (DE).  DE is postulated to have a negative pressure and is associated with the vacuum energy of space itself \cite{ref8,ref9}. This negative pressure leads to the repulsive gravitational effects that drive the universe's accelerated expansion \cite{ref10,ref11}. Current theories suggest that dark energy constitutes about $68 \%$ of the total energy content of the universe. This is the dominant contributor to the total matter of the universe, along with about $27 \%$ of dark matter, and the rest in the form of normal matter (atoms).\\

To explain the cosmic expansion, a term has been introduced in Einstein’s equation as an alternative to DE. This term causing space to repel itself and leading to the observed accelerated expansion is called a cosmological constant. However, this approach faces two major challenges: the problems of fine-tuning and cosmic coincidence. To address these challenges, alternative theories of gravitation have been explored. Modified theories of gravity involve modifying the gravitational part of Einstein's equations in various ways. These modifications could potentially explain the accelerated expansion. Such theories might provide explanations for the accelerated expansion by altering fundamental laws of gravity rather than invoking a new form of energy. Several theories have been developed through modifications of Lagrangian and curvature of Einstein’s equations in general relativity \cite{ref12,ref13,ref14,ref15,ref16,ref17,ref18,ref19,ref20}. Among these modified theories, $f(R)$ theory is one of the appropriate alternatives of GR theory, which faces challenges in passing certain observational tests \cite{ref21,ref22,ref23,ref24,ref25, ref26,ref27}. The $f(R,T)$ theory is another modified version of $f(R)$ theory that explains the accelerated expansion of the universe at late times \cite{ ref28,ref29,ref30,ref31,ref32}. \\

``$f(Q,T)$ gravity" is an extension of $f(Q)$ gravity within the field of theoretical physics and cosmology that extends and modifies Einstein's General Theory of Relativity (GR). In General Relativity, the gravitational interaction between masses is described by the curvature of space-time caused by the distribution of matter and energy. In $f(Q,T)$ theory the energy-momentum tensor describes the distribution of matter and energy in space-time, and the quadratic scalar is a mathematical term derived from the geometry of the space-time manifold. By allowing for a wider range of functions and interactions between these two terms, $f(Q,T)$ gravity aims to provide a more comprehensive and flexible description of gravitational phenomena. In some $f(Q,T)$ gravity models, modifications to the gravitational field equations can lead to predictions that mimic the effects of dark matter and dark energy, thus providing alternative explanations for the observed cosmic acceleration and galactic rotation curves. Researchers and cosmologists are actively exploring various $f(Q,T)$ gravity models to better understand their implications for cosmology, astrophysics, and fundamental physics \cite{ref33,ref34,ref35}.  In the study of the expansion of the current universe, Bhattacharjee et al. \cite{ref36}, and Arora et al. \cite{ref37} have also focused on the framework of $f(Q,T)$ gravity. This approach allows researchers to explore modifications to the standard General Relativity equations and their implications for the evolution of the cosmos. Zia et. Al. \cite{ref38} discusses the general form of the transient behavior of $f(Q,T)$ gravity. This likely includes a study of how this modified gravity theory behaves over time, especially in comparison to standard General Relativity. The reference \cite{ref39} appears to focus on explaining the parameters of a linear case model within the context of $f(Q,T)$ gravity. This kind of study helps establish the relationships between the theory's parameters and its predictions.\\

Weyl-type $f(Q,T)$ gravity is a fascinating extension of gravitational theory that introduces a novel approach to understanding the fundamental forces governing the universe. In this framework, the traditional Einstein-Hilbert action is modified by incorporating a function of the Weyl tensor ($Q$) and the trace of the energy-momentum tensor ($T$). This modification allows for a richer description of gravity and space-time geometry. In gravitational physics, the Weyl tensor represents the traceless part of the Riemann curvature tensor. It characterizes the gravitational tidal forces and describes the intrinsic geometry of space-time. The energy-momentum tensor encodes information about the distribution of matter and energy in space-time. Weyl-type $f(Q,T)$ gravity is a rapidly evolving field of research. Scientists Yixin et al.  \cite{ref40}, Yang et al. \cite{ref41}, Gadbail et al.  \cite{ref42} are actively investigating various models, cosmological implications, and observational tests to validate or refine this framework. In the context of Weyl-type $f(Q,T)$ gravity, reference \cite{ref43} discusses the impact of viscosity on cosmological evolution. This suggests an investigation into how the presence of viscosity affects the dynamics of the universe within the framework of this modified gravity theory. Gadbail et al.  \cite{ref44} have used parametrization of DP to explore the cosmological dynamics of the universe in Weyl-type $f(Q,T)$ gravity. \\

In the present study, we have explored the cosmological dynamics of the universe in  Weyl-type $f(Q,T)$ by assuming the parametrization $H(z) = H_0 (1+z)^{1+ \alpha + \beta} exp \left(\frac{- \beta z}{1+z} \right)$. The present study is organized as: 
In Section $2$, we have discuss the formulation in Weyl-type $f(Q,T)$ gravity. Using parametrization of the deceleration parameter, the solution of field equations has been proposed in Section $3$.  In Section $4$, we have discussed the observational constraints with Hubble data points in the range $0\leq z \leq 2.36$. Some features of the proposed model are explained in Section $5$. The viability of the derived model is also examined through $w_D-w_D'$ in Section $6$. The quintessence field is defined in Section $7$. State-finders and sound speed are explained in Sections $8$ and $9$. Classical linear and nonlinear energy conditions are defined in Section $10$. Finally, in Section $11$  conclusion is summaries.

\section{ Field equation of Weyl type $f(Q, T)$ gravity}
Weyl proposed a new geometry in 1918, that introduces a relationship characterized by the change in both the orientation and magnitude of a vector under parallel vector transport.  The formulation of the action in Weyl gravity is defined as \cite{ref40}.
\begin{equation}
\label{1}
S=  
\int {\sqrt{-g}\left[-\frac{1}{4} W_{\mu v} W^{\mu \nu}-\frac{1}{2} M^2 w_\mu w^\mu + \kappa^2 f(Q, T)+ \left(R+6 \nabla_a w^\alpha - 6 w_\alpha w^{\alpha}\right) \lambda+\mathcal{L}_m\right] d^4 x},
\end{equation}
with $16 \pi G\kappa^2= 1$. Non-metricity Q is an essential aspect of our theory that can be determined through the process. 
\begin{equation}
\label{2}
Q \equiv-g^{\mu v}\left(L_{\beta \mu}^\alpha L_{\nu \alpha}^\beta-L_{\beta \alpha}^\alpha L_{\mu \nu}^\beta\right)
\end{equation}
here $L_{\mu \nu}^\lambda$ is define as
\begin{equation}
\label{3}
L_{\mu \nu}^\lambda= \frac{- g^{\lambda \gamma}}{2} \left(Q_{\mu \gamma \nu}+Q_{\nu \gamma \mu}-Q_{\gamma \mu \nu}\right) .
\end{equation}
The covariant derivative of metric in the Riemannian geometry is zero but in Weyl geometry, non metricity tensor is defined as
\begin{equation}
\label{4}
-\widetilde{\Gamma}_{\alpha \mu}^\rho g_{\rho \nu}-\widetilde{\Gamma}_{\alpha \nu}^\rho g_{\rho \mu} + \partial_\alpha g_{\mu \nu} =   \tilde{\nabla}_\alpha g_{\mu \nu} \equiv Q_{\alpha \mu \nu}.
\end{equation}
From Eq. (3) and (4), we find the following relation 
\begin{equation}
\label{5}
Q=-6 \omega^2 .
\end{equation}
The generalized equation by variation on Eq.(1) is
\begin{equation}
\label{6}
 -\omega_\mu \left(M^2+12 \kappa^2 f_Q+12 \lambda\right)  + \nabla^\nu W_{\mu \nu} = 6  \lambda  \nabla_\mu.
\end{equation}
The effective mass of the vector field is
\begin{equation}
\label{7}
M_{e f f}^2=M^2+12 \kappa^2 f_Q+12 \lambda  .
\end{equation}
The variation on action (1) w.r.t. metric tensor and weyl vector yield
$$
\left(\frac{T_{\mu \nu}+S_{\mu \nu}}{2}\right)-\kappa^2 f_T\left(T_{\mu \nu}+\Theta_{\mu \nu}\right) = -\frac{\kappa^2}{2} g_{\mu \nu} f -6 k^2 f_Q \omega_\mu \omega_\nu+ \left(-6 \omega_\mu \omega_\nu+ R_{\mu \nu} +3 g_{\mu \nu} \nabla_\rho \omega^\rho\right) \lambda
$$
\begin{equation}
\label{8}
+ g_{\mu \nu} \square \lambda -6 \omega_{(\mu} \nabla_{v)} \lambda +3 g_{\mu \nu} \omega^\rho \nabla_\rho \lambda  -\nabla_\nu \lambda \nabla_\mu .
\end{equation}
here, $f_Q$ and $f_T$ are the partial derivatives of $f(Q,T)$ w.r.to $Q$ and $T$ respectively.
$T_{\mu \nu}$ and $\Theta_{\mu v}$ are define as
%The definition of $T_{\mu \nu}$ and $\Theta_{\mu v}$ is
\begin{equation}
\label{10}
T_{\mu \nu} \equiv-\frac{2}{\sqrt{-g}} \frac{\delta\left(\sqrt{-g} L_m\right)}{\delta g^{\mu \nu}} .
\end{equation}
\begin{equation}
\label{11}
\Theta_{\mu \nu}=g^{\alpha \beta} \frac{\delta T_{\alpha \beta}}{\delta g_{\mu \nu}}=g_{\mu \nu} L_m-2 T_{\mu \nu}-2 g^{\alpha \beta} \frac{\delta^2 L_m}{\delta g^{\mu \nu} \delta g^{\alpha \beta}} .
\end{equation}
Here, Re-scaled energy momentum tensor $S_{\mu \nu}$ is given by
%Here, $S_{\mu \nu}$ is the re-scaled energy momentum tensor given by
\begin{equation}
\label{12}
S_{\mu \nu} = - \frac{g_{\mu \nu}}{4} \omega_{\rho \sigma} W^{\rho \sigma} +W_{\mu \rho} W_v^\rho - \frac{M^2}{2} g_{\mu \nu} \omega_\rho \omega^\rho+M^2 \omega_\mu \omega_\nu  ,
\end{equation}
and
\begin{equation}
\label{13}
W_{\mu \nu}=\nabla_\nu w_\mu-\nabla_\mu \omega_\nu .
\end{equation}
%It is also noted that the expression for the divergence of the matter 

The energy-momentum tensor in the weyl gravity is defined as 
\begin{equation}
\label{14}
\nabla^\mu T_{\mu \nu}=\frac{\kappa^2}{1+2 \kappa^2 f_T}\left[-f_T \nabla_\nu T-2 T_{\mu \nu} \nabla^\mu f_T + 2 \nabla_\nu\left(f_T \mathcal{L}_m\right)\right] .
\end{equation}
As a result, the matter energy-momentum tensor is not conserved, meaning that it does not remain constant over time. This non-conservation is indicated by the presence of a non-zero right-hand side in the equation. Physically, the non-conservation of the matter energy-momentum tensor implies the existence of an additional force acting on massive test particles. This force affects the motion of these particles, causing them to be non-geodesic. Geodesic motion refers to the path that a free particle would naturally follow in the absence of any external forces \cite{ref45}.

\section{Cosmological model with deceleration parameter}
For the purpose of modeling, we consider a spatially flat FLRW metric. 
\begin{equation}
\label{15}
d s^2=  \delta_{i j} d x^i d x^j a^2(t) -d t^2,
\end{equation}
here $a(t)$ represents the scale factor. Because of spatial symmetry,
the vector field $w_\mu$ can be assume as $w_\mu=[\psi(t), 0,0,0]$. Using this we get, 
$$
  Q =-6 \omega^2= 6 \psi^2(t), and~~ \omega^2= \omega_\mu \omega^\mu=-\psi^2(t)
$$
The energy momentum tensor for the perfect fluid is given by:
\begin{equation}
\label{16}
T_{\mu \nu}=(\rho+p) u_\mu u_\nu +p g_{\mu \nu} .
\end{equation}
here $\rho$ and $p$  are energy density and pressure, respectively. 
$u^\mu$ is the four velocity vector satisfying $u_\mu u^\mu=-1$. So we have 
$$
\Theta_\nu^\mu=\delta_\nu^\mu p-2 T_\nu^\mu=\operatorname{diag}(2\rho+p,-p,-p,-p), and~~
T_\nu^\mu=\operatorname{diag}(-\rho, p, p, p) .
$$
The generalized Proca equation and constraint of flat space in the cosmological case can be found as:
\begin{equation} \label{17}
\psi= \psi^2-3 H \psi + \dot{H} + 2 H^2, 
\end{equation}
\begin{equation}
\label{18}
\dot{\lambda}=\left(-\frac{1}{6} M^2-2 \kappa^2 f_Q-2 \lambda\right) \psi=-\frac{1}{6} M_{e f f}^2 \psi, 
\end{equation}
\begin{equation}
\label{19}
\partial_i \lambda=0 .
\end{equation}
Using equation (8) and (15), the generalized Friedmann equations define as,
\begin{equation}
\label{20}
\kappa^2 f_T(\rho+p)+\frac{1}{2} \rho=\frac{\kappa^2}{2} f- \psi^2 \left(6 \kappa^2 f_Q+\frac{1}{4} M^2\right)  
-\left(\psi^2-H^2\right) 3 \lambda - (\psi-H) 3 \dot{\lambda}.
\end{equation}
\begin{equation}
\label{21}
-\frac{1}{2} p= \left(2 \dot{H}+3 H^2+3 \psi^2\right) \lambda + \frac{ \kappa^2 f}{2} +\frac{M^2 }{4} \psi^2 +\ddot{\lambda} + \dot{\lambda} (2 H + 3 \psi).
\end{equation}
Eliminating the derivatives of $\lambda$ from Eqs. (17) (18), (19), and (20), we get following to expressions.
\begin{equation}
\label{22}
 \frac{1}{2}\left(1+2 \kappa^2 f_T\right) \rho+\kappa^2 f_T p=\frac{\kappa^2}{2} f+\frac{m^2 \psi^2}{4} 
 +3 \lambda\left(H^2+\psi^2\right)-\frac{1}{2} m_{e f f}^2 H \psi ,
 \end{equation}
 \begin{equation} \label{23}
\frac{1}{2}(p+\rho)\left(1+2 \kappa^2 f_T\right)=\frac{1}{6}\left(\dot{\psi}-\psi H  +\psi^2\right) m_{eff}^2 
+2 \kappa^2 f_Q \psi-2 \dot{H} \lambda  .
\end{equation}
where $f_Q$ $\&$ $f_T$ are the partial derivatives w.r.to $Q$ and $T$ and {dot}(.) indicates time derivative. In the present study, we assume $f(Q, T)=\delta Q+\frac{\gamma}{6 \kappa^2} T$;  here, $\delta$ and $\gamma$ are the parameters. $M^2=\frac{m^2}{\kappa^2}$ represents the Weyl field's mass and $\kappa^2$ denotes the strength of coupling between matter and the Weyl geometry. For this case, we assume $M=0.95$ \cite{ref40}. It is important to note that for $\gamma=0$ and $\alpha=-1$, Weyl type $f(Q,T)$reduces to  $-Q$ a successful reduction of GR. On the other hand, for $T=0$, it turns to $f(Q)=\alpha Q$, which is equivalent to GR and is consistent with cosmological assessments and observational findings. Additionally, by utilizing the expression $\bar{\nabla}_{\lambda}. g_{\mu \nu}=-\omega_{\lambda}. g_{\mu \nu}$ we can obtained $\psi(t)=-6 H(t)$.\\
Now, the following expressions of the pressure and energy density can be determined using Eqs. (21) and (22).
\begin{equation}
\label{23}
p=-\left( 36 \left(\frac{18}{\gamma+3}(\delta+1)+\frac{3 M^2}{2(\gamma+3)}\right)+\frac{18}{2 \gamma+3}\right)  H^2-\frac{(18 \gamma + 36) \dot{H}}{(2 \gamma+3)(\gamma+3)},
\end{equation}
and
\begin{equation}
\label{24}
\rho=\left(\frac{-(99 \gamma+ 216)(24 \delta +25)}{(4\gamma+8)(\gamma+3)}+\frac{29 \gamma+72}{(4 \gamma+6)(\gamma+2)}\right) H^2-\frac{9 \gamma \dot{H}}{(4 \gamma+6)(\gamma+3)} ,
\end{equation}
We obtain EoS parameter $\omega_{eff }=\frac{p}{\rho}$ from equations (23) and (24),
%From Eqs. (20) and (21), the EoS parameter $\omega_{eff }=\frac{p}{\rho}$ can be analytically expressed as,
\begin{equation}
\label{26}
\omega_{e f f}=\frac{-\left(36\left(\frac{18}{\gamma+3}(\delta+1)+\frac{3 M^2}{(2 \gamma+6)}\right)+\frac{18}{2 \gamma+3}\right) H^2-\frac{18(\gamma+2) \dot H}{(2 \gamma+3)(\gamma+3)}}{\left(\frac{-9(11 \gamma+24)(24 \delta + 25)}{(4 \gamma+ 8)(\gamma+3)}+\frac{29 \gamma+72}{2(2 \gamma+3)(\gamma+2)}\right) H^2-\frac{9 \gamma \dot{H}}{2(2 \gamma+3)(\gamma+3)}} .
\end{equation}
{\large{\bf Parametric form of Deceleration Parameter}}\\
The deceleration parameter, $q$, describes the rate of change of the expansion of the universe. It provides information about the universe's expansion at a particular point in its history. The deceleration parameter can be used to classify the evolution of the universe into different phases: $q > 0 $  indicates that the expansion of the universe is decelerating. In other words, the rate of expansion is slowing down over time. This is typical for a universe dominated by matter and radiation. $q = 0$ signifies that the universe's expansion is neither accelerating nor decelerating. It could correspond to a transitional phase between deceleration and acceleration. $q < 0$ suggests that the expansion of the universe is accelerating. In this scenario, the rate of expansion is increasing over time. This can be attributed to the presence of dark energy, a hypothetical form of energy with negative pressure that counteracts the gravitational attraction of matter.
The parametric form of Deceleration Parameter (DP) is considered as\cite{ref46}.

\begin{equation} \label{27}
q =  \beta z (1+z)^{-1} + \alpha 
\end{equation}
here, $\alpha$ and $\beta$ are fixed values. The current deceleration parameter value, denoted as $q_{0}$ equal to $\alpha$, at $z = 0$. Furthermore, for specific value of $\alpha = 1/2$ and $\beta = 0$, the value of deceleration parameter is equal to $\frac{1}{2}$ which indicate that the universe is dominated by dark matter. In terms of redshift $z$, the Hubble parameter can be defined as:  
\begin{equation}
\label{28}
H = - (z+1)^{-1} \frac{dz}{dt} 
\end{equation}
From Eqs. (26) and (27), we can explain the Hubble parameter as
\begin{equation}
\label{29}
H(z) = H_0 (z+ 1)^{\alpha + \beta +1} exp \left(\frac{- \beta z}{1+z} \right)
\end{equation}
Here, $H_0$ denotes the current value of the Hubble parameter.\\

\section{Observational data analysis}
In this segment, we have applied the Markov Chain Monte Carlo (MCMC) technique to find the best-fitted optimal parameter values of the model. Here, $\zeta_{th}$ signifies the theoretically predicted value, while $\zeta_{ob}$ stands for the observational value. The $\chi^2$ evaluation function is examined as follows:
\begin{equation}
\label{30}
\chi^{2}_{\zeta}\left(P \right)=\sum_{i=1}^{} {\frac{\left(\zeta_{th}(P)-\zeta_{ob}\right)^2}{\sigma_\zeta^2}}.
\end{equation}
In this context, $P$ signifies the parameters of the model, while $\sigma_\zeta$ denotes the standard errors associated with measurements of a physical measure. Here, the parameter vector is denoted by $P = (H_{0},\alpha, \beta)$. Through a statistical process that involves minimizing the estimation function $\chi^2$, it becomes possible to determine the most probable values of parameters. For this purpose, we utilized the datasets of experimental findings like the observational Hubble data (OHD) consisting of 57points within the range $0.07 \leq z \leq 2.36$ \cite{ref47,ref47a}, Pantheon dataset of 1048 Type Ia Supernovae (SN Ia) within the redshift range $0.01 \leq z \leq2.26$  \cite{ref48}, and an observation dataset related to baryon acoustic oscillation (BAO) \cite{ref49,ref50,ref51,ref52,ref53}.  The confidence contour plots with $1\sigma$ ($68 \%$) and $2\sigma$ ($95\%$) confidence in two dimensional for the given model are shown in Figure 1. The summary of best-estimated values of parameters for the derived model are tabulated in Table 1.
%%%%%%%%%%%%%%%%%%%%%%%%%%%%%%% Table 1 %%%%%%%%%%%%%%%%%%%%%%%%%%%%%%%%%%%%%%%%%%
\begin{table}[H]
	\caption{ The values obtained of parameters for different observational dataset }
	\begin{center}
		\begin{tabular}{|c|c|c|c|c|c|}
			\hline
			Parameters & $H_0$& $\alpha$ & $\beta$	   & $z_{t}$ & $q_0$  \\
			\hline
			BAO+OHD57+Pantheon & $71.17\pm 0.25$ & $-0.663\pm0.030$ & $1.488\pm0.087$   & $0.804^{+0.175}_{-0.132}$ & $-0.663\pm0.030$ \\
			\hline
			BAO+OHD57 & $70.0^{+1.4}_{-1.2}$  & $-0.638\pm 0.076$ & $1.33\pm 0.16$  & $0.922^{+0.644}_{-0.316}$ & $-0.638\pm 0.076$ \\
			\hline
			Pantheon & $71.14^{+0.32}_{-0.28}$ & $-0.676^{+0.060}_{-0.071}$ & $1.77^{+0.37}_{-0.31}$   & $0.618^{+0.429}_{-0.214}$ & $-0.676^{+0.060}_{-0.071}$ \\
			\hline
			OHD57 & $68.60\pm1.6$ & $-0.571\pm0.092$ & $1.34^{+0.19}_{-0.17}$  & $0.752^{+0.574}_{-0.290}$ & $-0.571\pm0.092$ \\ 
						
			\hline
		\end{tabular}
	\end{center}
\end{table} 
For the combined observed data set, the estimator $\chi^{2}_{total}$ can be expressed as
 \begin{equation}
  \label{31}
\chi^{2}_{total} = \chi^{2}_{OHD} + \chi^{2}_{Pantheon} + \chi^{2}_{BAO}
\end{equation}
\begin{figure}[H]
	\centering
	\includegraphics[scale=0.8]{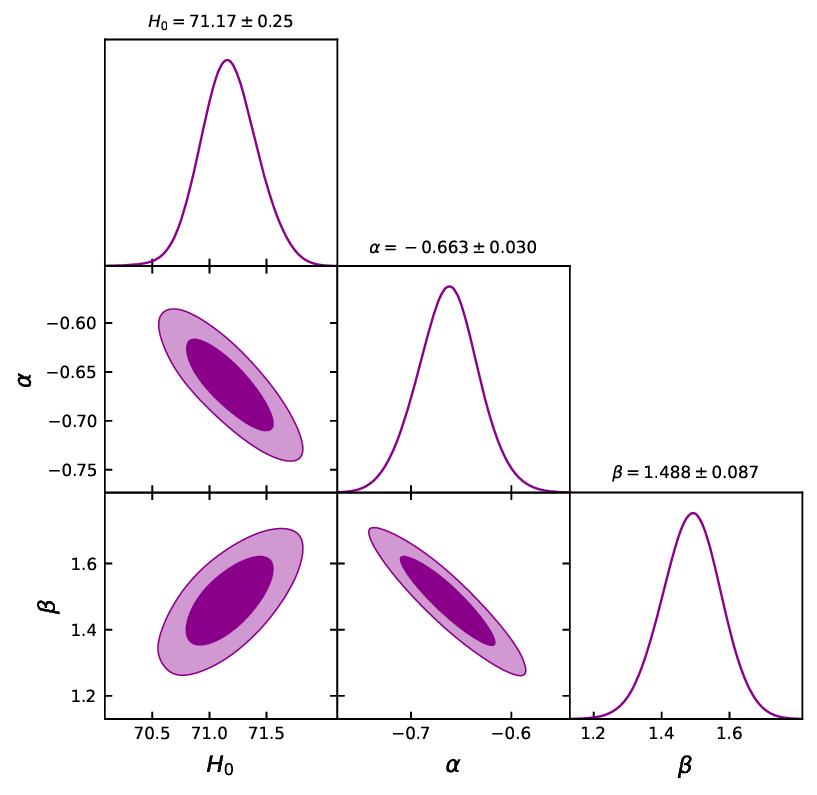} 
	\caption{ Confidence levels from the combination of OHD+Pantheon+BAO for model parameters. }
\end{figure}
For different sets of observations of BAO, OHD, and Pantheon, the best outcomes on Hubble tension $H_0$ and other model parameters for the proposed are found in nice agreement with recent findings \cite{ref47,ref53a,ref53b,ref53c,ref53d,ref53e,ref53f,ref53g}. 
\section{Features of the Model}
\begin{figure}[H]
	(a)\includegraphics[width=8cm,height=6cm,angle=0]{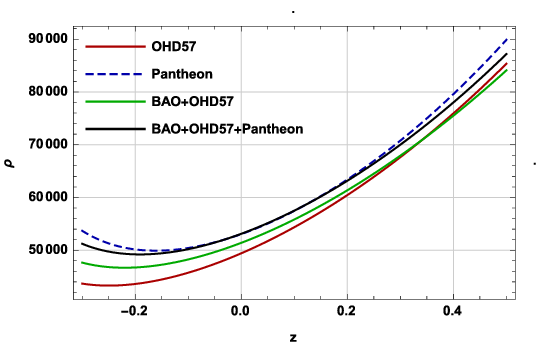}
    (b)\includegraphics[width=8cm,height=6cm,angle=0]{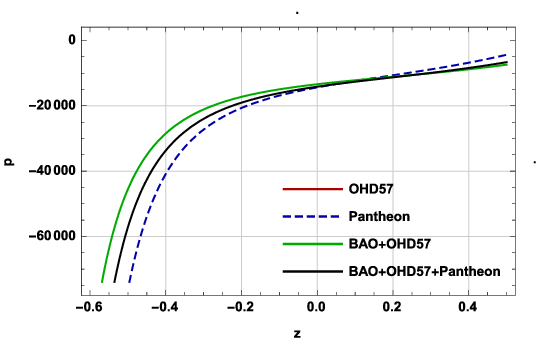}
\caption{(a) Energy density curve against $z$, (b) Plot of the cosmic pressure.}
\end{figure}
For all observational datasets energy density is positive of Weyl-type of $f(Q,T)$ gravity as shown in Figure 2(a). For various observational datasets, the behavior of cosmic pressure is demonstrated in Figure 2(b). It has been seen that the pressure is negative throughout the whole evolution of the universe. The negative behavior describes the present accelerated expansion of the universe. It also explains that with the universe's development, the energy density diminishes, leading to an increase in the volume of space. \\
\begin{figure}[H]
	\centering
	\includegraphics[width=8cm,height=6cm,angle=0]{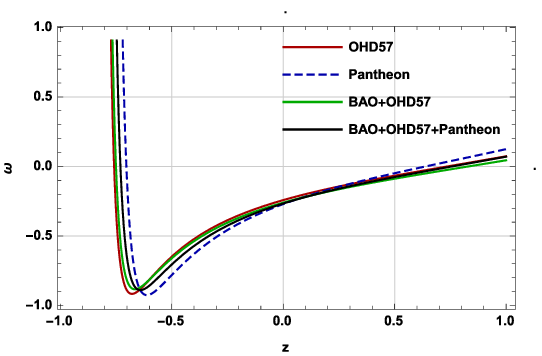}
	\caption{Plot of EoS parameter vs redshift $z$. }
\end{figure}
The equation of state parameter is a crucial parameter in cosmology and astrophysics that describes the relationship between the pressure and the energy density of a substance, such as matter or dark energy, in the universe. It is used to characterize the nature of the substance and its effect on the expansion of the universe. The value of the equation of state parameter can vary for different substances: For non-relativistic matter, the pressure is negligible compared to the energy density, so $\omega$ is close to 0. For relativistic particles, such as photons and neutrinos, the pressure is significant compared to the energy density, leading to $\omega = \frac{1}{3}$. This is due to the relativistic equation of state. The equation of state parameter for a cosmological constant is $\omega = -1$, which means that its pressure is negative and equal in magnitude to its energy density. This negative pressure is responsible for the accelerated expansion of the universe. Some theories propose that dark energy might not be a cosmological constant but instead a dynamic field evolving over time. In these cases, the EoS parameter can vary and be different from -1. It could be greater than -1 (indicating repulsive gravity) or less than -1 (indicating even more rapid expansion). From the figure, we can observed that the derived model initially lies in quintessence era and advanced to Chaplygin gas scenario in late time \cite{ref53a,ref53f}.

%%%%%%%%%%%%%%%%%%%%%%%%%%%%%%%%%%%%%%%%%%%%%%
\section{$w_D-w_D'$ Analysis}
Caldwell and Linder \cite{ref54} conducted a study to investigate the changing dynamics of quintessence models of dark energy in the phase plane represented by $w$ and its time derivative $w'$, where  $w'$ is the derivative of $w$ with respect to the logarithm of the scale factor $a$  i.e. ($w' = \frac{dw}{d ln(a)}$). They demonstrated that these models classified into two distinct regions in the phase plane, known as the `thawing' ($w_D <0, w'_D>0$) and `freezing' regions ($w_D <0, w'_D < 0$) with quite different behavior in the $w_D-w_D'$ plane \cite{ref55,ref56,ref57}. 
\begin{figure}[H]
	\centering
	\includegraphics[width=8cm,height=6cm,angle=0]{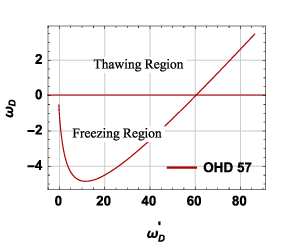}
	\includegraphics[width=8cm,height=6cm,angle=0]{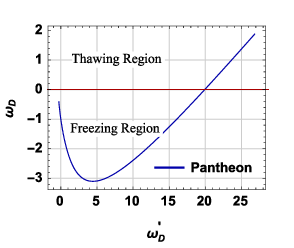}
	\includegraphics[width=8cm,height=6cm,angle=0]{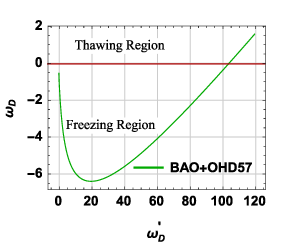}
	\includegraphics[width=8cm,height=6cm,angle=0]{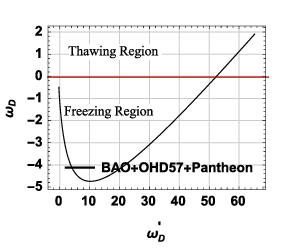}
	\caption{Plots of $w_D-w_D'$ for Weyl-type $f(Q,T)$ gravity vs $z$. }
\end{figure}
We have plotted $w_D'$ against $w_D$ for all the observational values in Figure 4 to construct the $w_D-w_D'$ plane. The curves in the plot exhibit both freezing and thawing regions for different observational datasets. We observe that our model's trajectories primarily diverge in the freezing region, as supported by observational data, indicating a more accelerated expansion of the universe in this area.

\section{Quintessence field in Weyl-type of $f(Q, T)$ gravity}
Quintessence is a theoretical concept in cosmology that refers to a hypothetical scalar field responsible for dark energy. It is considered one of the leading explanations for the accelerated expansion of the universe. The term "quintessence" comes from ancient cosmology, where it was used to describe the fifth element that completes the classical four elements of earth, water, air, and fire. In modern cosmology, quintessence is associated with a scalar field that has a positive energy density and negative pressure, causing it to have repulsive gravitational effects. This negative pressure leads to the expansion of the universe at an accelerating rate, counteracting the attractive gravitational force of matter and radiation. The presence of quintessence would help explain the observed phenomenon of DE. The dynamics of the quintessence field depend on its potential energy function. Different potential energy functions result in different behaviors of the quintessence field over cosmic time. The scalar potential is a basic idea in physics, which gives each point in space a unique value to describe the energy associated with a scalar field. The specific physical system under examination will determine the properties of this field and its potential. Since negative energy can lead to solutions that defy the laws of physics, a positive scalar potential is often linked with stable configurations in physics.
The negative scalar potential is demonstrated in specific frameworks like models involving dark energy or cosmological inflation. The behavior of a scalar potential is ultimately determined by physics and the particular values of relevant parameters.\\
The action for the quintessence field is explained as, \cite{ref58}
\begin{equation}
\label{31}
S=\int \sqrt{-g} d^4 x\left[-\frac{1}{2} g^{i j} \partial_i \phi \partial_j \phi-V(\phi)\right],
\end{equation}
where $g$ is the metric determinant $\&$  $g^{i j}$ is the metric in the above Eq.(31),  $V(\phi)$ is the potential for the quintessence field $\phi$. The energy density and pressure for the quintessence scalar field by varying the action w.r.t. the metric and $\phi$  are defined as:
\begin{equation}
\label{33}
 \rho_\phi=\frac{\dot{\phi}^2}{2}+V(\phi), \\
  p_\phi=\frac{\dot{\phi}^2}{2}-V(\phi) .
\end{equation}
\begin{figure}[H]
	(a)\includegraphics[width=8cm,height=6cm,angle=0]{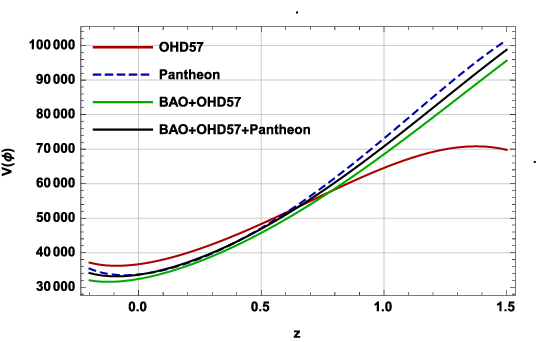}
	(b)\includegraphics[width=8cm,height=6cm,angle=0]{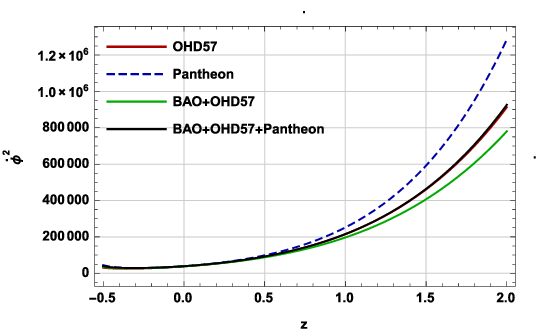}
     \caption{(a) Plot of scalar potential vs $z$ (b) Graph of scalar field $\phi $ vs $z$ for all four sets of best fitted values. }
\end{figure}
Figures 5(a) and 5(b) illustrate the changes in the scalar field $\phi$ and its corresponding potential $V(\phi)$ within the quintessence model as the redshift z varies. The plots were generated using the estimated values of parameters.
For these suitable choices of parameters, the field gets trapped in the local minimum because the kinetic energy during a scaling regime is small. The field then enters a regime of damped oscillations leading to an accelerating universe \cite{ref53f,ref58,ref58a}.

\section{statefinders}
Statefinder is a cosmological diagnostic tool used in astrophysics and cosmology to distinguish between different dark energy (DE) models. It has been computed for various existing models of dark energy, that provide an effect between different forms of dark energy in the cosmological plane. This plane represents distinct, well-known regions of the universe. For instance, the pair $(s>0$ and $r<1)$ corresponds to the region of quintessence DE eras, $(r, s)=(1,1)$ represents the CDM limit, $((r, s)=(1,0))$ signifies the $\Lambda$CDM limit, and $(s<0$ and $r>1$ ) indicates the Chaplygin gas region. Using this diagnostic tool, one can assess how closely a dark energy model resembles $\Lambda$CDM dynamics \cite{ref59,ref60}. Alam et al. \cite{ref61} have defined the state-finders ($r$, $s$) as following
 
\begin{equation}
\label{34}
r=\frac{\dddot{a}}{a H^3}, s=\frac{r-1}{3\left(q-\frac{1}{2}\right)}
\end{equation}
where $a$ is the scale factor of the universe, H is the Hubble parameter, and the primes represent derivatives with respect to the cosmic time. State-finder depends on the scale factor and its time derivative. The state-finders can also read as
\begin{equation}
\label{35}
r=2 q^2+q-\frac{\dot{q}}{H} \quad, s=\frac{2}{3}(q+1)-\frac{\dot{q}}{3 H\left(q-\frac{1}{2}\right)}
\end{equation}

\begin{figure}[H]
	\centering
	\includegraphics[width=8cm,height=6cm,angle=0]{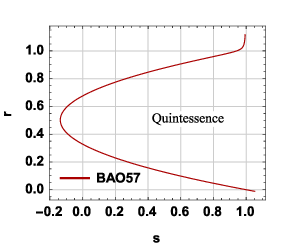}
	\includegraphics[width=8cm,height=6cm,angle=0]{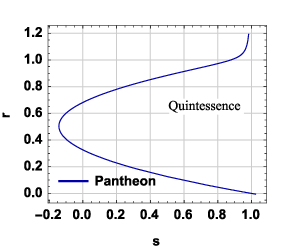}
	\includegraphics[width=8cm,height=6cm,angle=0]{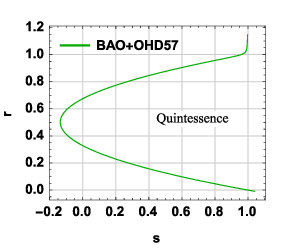}
	\includegraphics[width=8cm,height=6cm,angle=0]{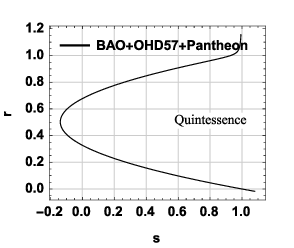}
	\caption{  Plots of $r$ versus $s$ for Weyl-type f(Q,T) theory. }
\end{figure}
For various best-fitted values of free parameters for the different observational datasets, the trajectories depicting the evolutionary behavior of the model in $r-s$ plane are shown in Figure 6. The figure shows that the model lies in the quintessence region $(r< 1, s>0)$. By using these Statefinder parameters, researchers can probe the dynamics of the universe and differentiate between different dark energy models more effectively \cite{ref53f,ref58,ref58a}.

\section{Speed of sound}
The sound speed $v^{2}_s$ should be necessarily less than the speed of light ($c$). The velocity of sound exists within the range $0 \leq v^{2}_s \leq 1$ with cosmic time, as we are working with gravitational units with a unit speed of light time \cite{ref62,ref63,ref64}.  The formula for the square of sound speed is:
\begin{equation}
\label{36}
v^2_{s} = \frac{dp}{d \rho}
\end{equation}
\begin{figure}[H]
	\centering
	\includegraphics[width=8cm,height=6cm,angle=0]{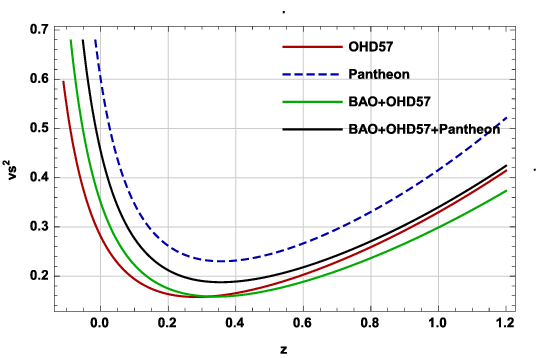}
	\caption{ Sound speed trajectory vs $z$. }
\end{figure}
To ensure the stability of the given theoretical model, the velocity ($v_s^2$) should lie within the range of -1 to 1 for the redshift scale. By examining Figure 7, we can see that the sound speed consistently lies within this specified range throughout the evolution of the universe. 
\section{Classical linear and nonlinear Energy Conditions}
In the context of general relativity, energy conditions are used to explore the properties and behaviors of space-time, and they play a role in various theorems and conjectures related to the nature of gravity and the possibility of constructing "exotic" configurations of matter that might lead to violations of certain physical principles. There are several energy conditions, and they fall into two main categories: linear energy conditions and non-linear energy conditions. In classical physics, energy conditions are linear, meaning that they have straightforward requirements on the stress-energy tensor that ensure the conservation of energy and other fundamental physical properties. However, in certain cases, nonlinear energy conditions are considered to explore more exotic scenarios that could potentially violate some of the standard assumptions. These conditions are used in the study of general relativity and are closely related to the concept of exotic matter and the possibilities of faster-than-light travel and traversable wormholes.\\

The linear energy conditions (ECs) within the framework of GR are mathematically explained as (1) Weak Energy Condition (WEC):  $\rho +  p \geq 0$ , $\rho \geq 0$, (2) Null Energy Condition (NEC):  $\rho + p \geq 0$, (3) Dominant Energy Condition (DEC):  $\rho - p \geq 0$ , (4) Strong Energy Condition (SEC):  $\rho + 3 p \geq 0$ \cite{ref65,ref66}. \\
In addition, non linear energy conditions are :
1.  The flux EC: $\rho^2 \geq p^2 $
2. The determinant EC: $\rho \Pi p_i \geq 0$
3. The trace of square EC: $\rho^2 + \sum p^2 \geq 0$ 

\begin{figure}[H]
	(a)\includegraphics[width=8cm,height=6cm,angle=0]{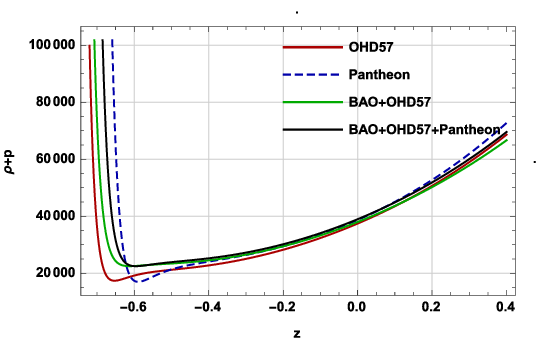}
	(b)\includegraphics[width=8cm,height=6cm,angle=0]{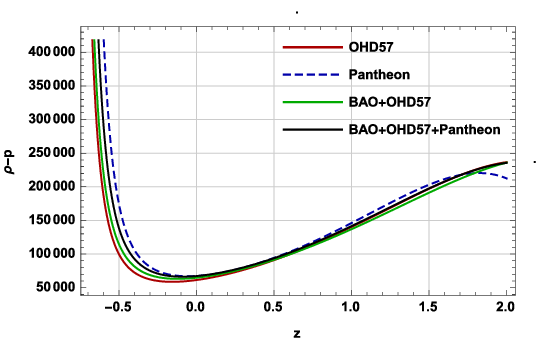}
	(c)\includegraphics[width=8cm,height=6cm,angle=0]{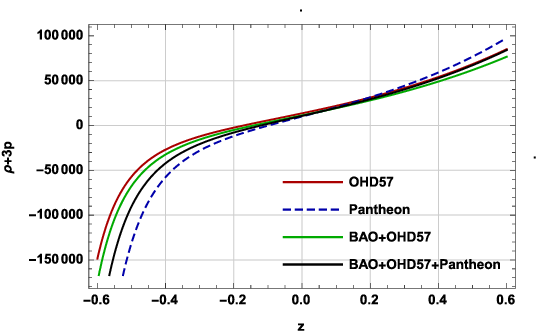}
	(d)\includegraphics[width=8cm,height=6cm,angle=0]{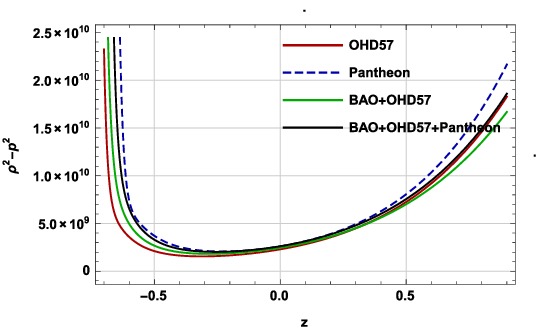}
	(e)\includegraphics[width=8cm,height=6cm,angle=0]{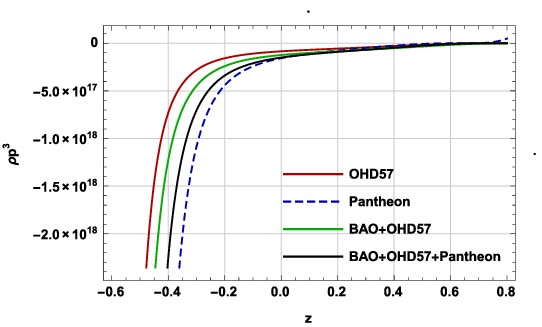}
	(f)\includegraphics[width=8cm,height=6cm,angle=0]{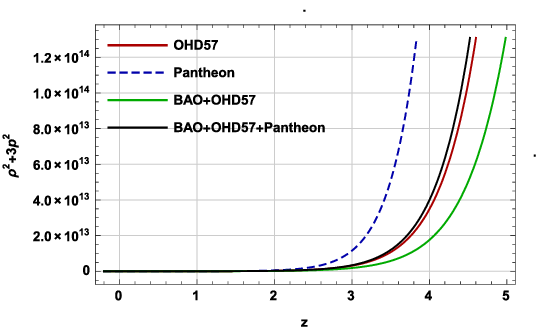}
	\caption{ Plots of classical linear and non-linear Energy conditions. }
\end{figure}
Figure 8 illustrates the accomplishment of energy conditions for the proposed model using the joint observational data from BAO, OHD, and Pantheon. The derived model satisfies WEC, DEC, and NEC, but violates SEC as depicted in Figure 8. The violation of SEC for the proposed model depicts the accelerated expansion of the cosmos and confirms the presence of exotic matter in the cosmos  \cite{ref65,ref66,ref67}. The term "flux" generally refers to the flow of some physical quantity through a surface. The tensor of stress energy is efficient in describing the flow of energy, momentum, or stress across a particular area. In the case of the Null Energy Condition, the NEC requires that the flux of the stress-energy tensor's components should be non-negative for all null (light-like) vectors. This helps to ensure that the energy density along any null geodesic remains non-negative. \\
 
The `trace' of the `stress-energy tensor' is the sum of its diagonal components, that typically correspond to pressure and energy density. In the context of the Strong Energy Condition, the trace condition often states that the summation of three times the pressure and the energy density should be non-negative. Figures 8(d), 8(e), 8(f) show the non-linear energy conditions. The Flux and the trace of the square of both ECs are satisfied. Non-linear energy conditions are particularly important in investigations of exotic matter and space-time geometries that might allow for phenomena like faster-than-light travel, time travel, or other violations of classical energy conditions. They help to identify regions of space-time where such exotic behavior might occur or be ruled out.

%%%%%%%%%%%%%%%%%%%%%%%%%%%%%%%%%%%%%%%%%%%%%%%%%%%%%%%%
\section{Concluding remarks}

This article focuses on examining a modified theory known as the Weyl $f(Q,T)$ gravity, where the relationship between metric and the Weyl vector plays an important role in determining the metric tensor’s  covariant divergence. Consequently, the geometrical properties of the theory are impacted by both the metric tensor and the Weyl vector. We have explored the universe's dynamics utilizing the parametric form of DP represented as $q(z) = \alpha + \frac{\beta z}{1+z}$. To estimate the free parameters $\beta$, $\alpha$, and $H_0$, we have utilized the latest experimental datasets of BAO, OHD, and Pantheon and implementing the application of the MCMC method. We have employed the open-source Python package `emcee’ for this purpose. Additionally, we have determined a few kinetic properties such as $w_D - w'_D$, quintessence, state-finders, sound speed, and energy conditions which provide further insights into the behavior and implications within this context. The highlights of the model are given below.
\begin{itemize}
\item
We have used three distinct sets of observational data: Baryon Acoustic oscillation(BAO), Observational Hubble Data (OHD), and data from Pantheon compilation. Implementing the MCMC statistical approach, we have estimated the free parameters $\beta$, $\alpha$, and $H_0$. Taking the joint dataset into account, the confidence contour for model parameters is plotted in Figure 1. The constrained values of model parameters are tabulated in Table 1.\\
\item
The evolutionary behaviours of the energy density and cosmic pressure are represented in Figures 2(a) and 2(b) respectively. For all observational datasets, the nature of the energy density is positive while nature of cosmic pressure is negative throughout the evolution.
\item
Figure 3 shows the EoS parameter $\omega$ for the three models of Weyl type $f(Q,T)$ gravity. It has been plotted for the observational values. During its evolution, the derived model initially lies in quintessence era and advanced to Chaplygin gas scenario in late time. 
\item
In Figure 4, we have described the $w_D-w_D'$ for all the observational values. It is found that the curve lies in both thawing and freezing regions.

\item 
The nature of scalar potential $V(\phi)$ and scalar field $\phi$ for Weyl type of $f(Q,T)$ gravity is depicted in Figure 5. We observe that the potential for the quintessence model is a decreasing function of $z$, which gives rise to an accelerated expansion. The behavior of the state-finders is shown in Figure 6. The trajectories for the proposed models lie in quintessence region $(r<1, s>0)$ as shown in the figure. For the proposed model, the sound speed’s nature can provide insights into its stability and characteristics. The trajectories of sound speed for different observational data sets are plotted in Figure 7. 
\item
Figure 8 shows linear and nonlinear energy conditions of the proposed model for observational datasets. The violation of SEC for the derived model depicts an accelerated expansion of the cosmos. Figures 8(d), 8(e), 8(f) show the non-linear energy conditions. The Flux and trace of square both ECs are satisfied.
\end{itemize}

In the manuscript, we have present a comprehensive study of an accelerated expanding cosmological model in Weyl $f(Q,T)$ gravity with the aid of a specific parametric approach.

%******************************************************************

\end{document}